\documentclass[fleqn,usenatbib]{mnras}

\usepackage{newtxtext,newtxmath}

\usepackage[T1]{fontenc}
\usepackage{ae,aecompl}
\usepackage{multirow}

\usepackage{graphicx}	
\usepackage{caption} 

\usepackage{cancel}     
\usepackage{color}      
\usepackage{ulem}       

\usepackage{float}
\usepackage[active, generate=file, extract-env={equation,align}]{extract}

\restylefloat{table}

\newcommand{\apspr}{APSPR}  

\newcommand{\erosita}{\textit{eROSITA}}
\newcommand{\erass}{\textit{eRASS}}
\newcommand{\startrack}{\texttt{StarTrack}}
\newcommand{\ulxlc}{\texttt{ULXLC}}

\newcommand{\ergss}{\mathrm{erg} \ \mathrm{s}^{-1}}

\newcommand{\csum}{\sum_{c=1}^{8}}

\newcommand{\Nsys}{N_{\rm sys}}       	 					

\newcommand{\Nalive}{N_{\rm A}}
\newcommand{\Ntransient}{N_{\rm T}}
\newcommand{\Ndead}{N_{\rm Hid}}

\newcommand{\Nulx}{N_{\rm ulx}(c)}        					
\newcommand{\Nnotulx}{\lnot \Nulx}  					
\newcommand{\Nnew}{N_{\rm new}(c)}        					
\newcommand{\Ndip}{N_{\rm dip}(c)}        					
\newcommand{\Ndeltaulx}{\Delta \Nulx}  				
\newcommand{\Ntransients}{N_{\rm T}(c)} 						
\newcommand{\Nalives}{N_{\rm A}(c)} 							
\newcommand{\Nobs}{N_{\rm obs}(c)} 		                    

\newcommand{\Nobscum}{\csum \Nobs}							
\newcommand{\Nalivecum}{\csum \Nalives}
\newcommand{\percobssamp}{\Nobscum / \Nsys} 					
\newcommand{\percobsdet}{\Nobscum / (\Nalive + \Ntransient)} 	
\newcommand{\perctransdet}{\Nobscum / \Ntransient}

\newcommand{\dimax}{\Delta i_{\mathrm{max}}}

\title{The impact of precession on the observed population of ULXs}

\author[N. Khan et al.]{Norman Khan,$^{1}$\thanks{E-mail: nk7g14@soton.ac.uk}
Matthew. J. Middleton,$^{1}$ Grzegorz Wiktorowicz,$^{2,3,4}$ \newauthor Thomas Dauser,$^{5}$ Timothy P. Roberts,$^{6}$ and Joern Wilms,$^{5}$\\
\\
$^{1}$School of Physics \& Astronomy, University of Southampton, Southampton, Southampton SO17 1BJ, UK\\
$^{2}$National Astronomical Observatories, Chinese Academy of Sciences, Beijing 100012, China\\
$^{3}$School of Astronomy \& Space Science, University of the Chinese Academy of Sciences, Beijing 100012, China\\
$^{4}$Nicolaus Copernicus Astronomical Center, Polish Academy of Sciences, Bartycka 18, 00-716 Warsaw, Poland\\
$^{5}$Remeis Observatory \& ECAP, Universit\"{a}t Erlangen-N\"{u}rnberg, Sternwartstr. 7, 96049 Bamberg, Germany\\
$^{6}$Centre for Extragalactic Astronomy \& Department of Physics, Durham University, South Road, Durham DH1 3LE, UK\\
}

\date{Accepted XXX. Received YYY; in original form ZZZ}

\pubyear{2021}

\begin{document}
\label{firstpage}
\pagerange{\pageref{firstpage}--\pageref{lastpage}}
\maketitle

\begin{abstract}
The discovery of neutron stars powering several ultraluminous X-ray sources
(ULXs) raises important questions about the nature of the underlying
population. In this paper we build on previous work studying simulated
populations by incorporating a model where the emission originates from a
precessing, geometrically beamed wind-cone, created by a super-critical inflow.
We obtain estimates -- independent of the prescription for the precession period of the wind --
for the relative number of ULXs that are potentially visible (persistent or
transient) for a range of underlying factors such as the relative abundance of
black holes or neutron stars within the population, maximum precessional angle,
and LMXB duty cycle. We make initial comparisons to existing data using a catalogue compiled
from \textit{XMM-Newton}.  Finally, based on estimates for
the precession period, we
determine how the \erosita \ all-sky survey (\erass) will be able to constrain
the underlying demographic.

\end{abstract}




\section{Introduction}

Ultraluminous X-ray sources (ULXs) are defined as extra-galactic, off-nuclear
point sources, with inferred isotropic luminosities in excess of $L \approx 1
\times 10^{39} \ \ergss$, around the Eddington limit for a typical stellar mass
black hole (see \citealt{Roberts_2007_ulx, Kareet_2017_review}). Originally
considered to be candidates for hosting intermediate mass black holes with
sub-Eddington accretion rates \citep{Colbert_1999_IMBHs}, it is now accepted
that \textit{most} ULXs contain stellar mass black holes (BH) or neutron stars
(NS) accreting at super Eddington (or `super-critical') rates.
To-date,
there have been ten ULXs discovered to harbour neutron star accretors, eight
identified through the detection of coherent pulsations (e.g.
\citealt{,NSULX_Bachetti_2014, NSULX_Furst_2016, Israel_2017_NSULX_5907, Tsygankov_2017_NSULX_SMCX3, Doroshenko_2018_NSULX_Swift_J0243, Carpano_2018_NSULX_NGC300, Sathyaprakash_2019_1313X2, 2020_Rodriguez_ApJ...895...60R}), and one via a cyclotron resonance scattering
feature (CRSF), potentially indicating the presence of strong quadrupole
magnetic fields (\citealt{2018_Brightman_CRSF, Middleton_Brightman_2019_M51};
note also the potential CRSF indirectly located in NGC 300 ULX-1 by
\citealt{Walton_2018_CRSF}, see also \citealt{Koliopanos_NGC300_CRSF}).

In the case of super-critical accretion, the accretion rate in Eddington units,
$\dot{m_{0}} = \dot{m} / \dot{m}_{ \rm Edd } > 1$, with the accretion flow first reaching the local Eddington limit around the spherization radius at $r_{\rm sph} \approx \dot{m_{0}}
r_{\rm in}$, where $r_{\rm in}$ is the inner radius of the disc, presumed to be
the innermost stable circular orbit (ISCO). In the case of a magnetised neutron star, as long as $r_{\rm sph}$ is
larger than the magnetospheric radius, then the Eddington limit is expected to
be reached locally in the disc of both NS and BH ULXs (conversely, for very
strong dipole fields, the flow will change accordingly - see
\citealt{Mushtukov_2017_PULXs_as_magnetars, Mushtukov_2018_mgntrs_high_acc_rate}).
At $r_{\rm sph}$, the radiation pressure inflates the disc towards scale heights
of order unity \citep{Poutanen_2007_ln}. In order to stay locally below the
Eddington limit, mass must be lost in the form of an outflow, which forms an
optically thick wind-cone (see \citealt{Poutanen_2007_ln}) which can collimate the
radiation from within. This `geometrical beaming' of the radiation naturally leads to
deviations from isotropy and a higher inferred luminosity
\citep{2001_King_ApJ...552L.109K}.

Under the assumption that geometrical beaming acts to some extent across the
entire ULX population  (i.e. ignoring the presence of very strong magnetic
fields - see \citealt{no_magnetars_in_ulxs} but also \citealt{Mush2021}), the
proportion of neutron stars and black holes within the ULX population has been
analytically estimated by \cite{Middleton_2017_demographics_from_beaming},
while estimates leveraging binary population synthesis have also recently been
explored \citep{Wiktorowicz_2019_obs_vs_tot}. Both studies predict that, whilst
NS systems almost certainly dominate the entire {\it intrinsic} population of
ULXs, {\it observationally} the populations of NS and BH ULXs may be
comparable (particularly for host regions with low metallicity), although this may be in conflict with spectral similarities between
the brightest ULXs (typically $>$ a few $\times$ 10$^{39} \ \ergss$) and those
systems confirmed to harbour neutron stars \citep{Pinto_2017AN....338..234P,
Walton_2018_CRSF}.

The light curves of several ULXs show modulations on month timescales, such as
the $\sim \mathrm{62}$ and $\sim \mathrm{55}$ day periods seen in both M82 X-1
and X-2 \citep{Kaaret_2006_Modulation_M82, Kong_2016_Modulation_M82_X2}, and the
$\sim \mathrm{78}$ day modulation detected in NGC 5907 ULX-1
\citep{Walton_2016_Modulation_5907}. These modulations may be explained by the
forced rotation of the accretion curtain in the case of a very high dipole
field NS (see \citealt{Mushtukov_2017_PULXs_as_magnetars}) or,
alternatively as suggested by \citet{Pasham_M82_precessing_disc} in the case of
M82 X-1, a precessing accretion disc. The precession of an accretion disc may
be driven by a variety of external torques, including tidal effects, radiation
pressure driven instabilities, Lense-Thirring precession, magnetic warping and
free-free precession  \citep{Fragile_2007_tilted_disc,
Maloney_1997_precessing_disc, Maloney_1998_warping_disc,
Pringle_1996_warping_disc, 2013_Lei_ApJ...762...98L}. Lense-Thirring
(solid-body) precession of the large scale-height disc and wind has recently
been proposed as the driving mechanism for the modulations (\citealt{Middleton_2018_Lense_Thirring,
Middleton_2019_Accretion_plane}) and is somewhat compelling as it requires a
misaligned spin and binary axis, the same requirement for the detection of
pulsations \citep{King_Lasota_2020_PULX_iceberg_emerges}. 

Under the assumption that ULXs are geometrically beamed sources with a
precessing disc/wind (regardless of the mechanism), then it follows that the
true population of ULXs is composed of (i) sources where we always view at low
inclinations to the wind-cone such that they are persistently {\it above} the
$1\times10^{39} \ \ergss$ limit, (ii) sources where we always view at high
inclinations to the wind-cone such that they are persistently {\it below} the
$1\times10^{39} \ \ergss$ limit (e.g. SS433, \citealt{Fabrika_2004_SS433};
\citealt{SS433_2021}), and (iii) sources which precess, such that the effective
observer inclination transitions between (i) and (ii)
\citep{2015_Middleton_MNRAS.454.3134M}. In this paper we investigate the effect
of geometrical beaming combined with precession on the observed population of
ULXs.

The recent launch of the \erosita \ mission \citep{eROSITAonSRG,
2021_Predehl_A&A...647A...1P} and the start of its all sky survey (\erass) will enable
the long-term X-ray variability of sources across the
entire sky to be probed. As we will demonstrate, the rate of discovery of ULXs in \erass \
monitoring may help in answering broad questions relating to the abundance of
BH and NSs in the ULX population. 

\section{Simulation Methods}
\subsection{Population Synthesis} \label{sec:population_synthesis}

Following the work of \citet{Wiktorowicz_2017_Origin} we obtained a sample of
simulated binary systems using the population synthesis code \startrack \
\citep{2008_Belczynski_Startrack, 2020_Belczynski_A&A...636A.104B}. The code
simulates the evolution of binaries while accounting for all processes that can
be important for the formation and evolution of ULXs such as the common
envelope phase, Roche Lobe overflow (RLOF) and tidal interactions. In addition, population synthesis 
invokes multiple formation channels in a variety of
stellar environments (metallicity, star formation history, age, etc.), and thereby provides synthetic data for comparison to 
observations. The
code outputs comprehensive information about system parameters, which we use to
calculate additional quantities required for our analysis.

\subsection{Luminosity and Beaming Factor} \label{sec:lum_b_fac}

We proceeded to select only those binary systems undergoing mass transfer from the
simulated sample; this provided 104,883 unique binary systems. For each of
these systems, their Eddington luminosity at each time interval was calculated
using $L_{\rm Edd} = 1.3 \times 10 ^{38} m \ \ergss$ (i.e. assuming a Hydrogen
composition), where $m$ is the compact object mass in solar units. The
Eddington mass transfer rate was calculated from $\dot{m}_{\rm Edd} = L_{\rm
Edd}/\eta c^{2}$ where we use $\eta \approx 0.08$ for both NS and BH systems.
Following \cite{Shakura_1973} and \cite{Poutanen_2007_ln}, we obtain the
intrinsic isotropic luminosity of the source without beaming $\mathrm{L_{\rm
iso}}$ (and ignoring energy lost in driving a wind, or advected in the case of
BHs):
    
\begin{equation}
L_{\rm iso} \approx
\left\{
	\begin{array}{ll}
		L_{\rm Edd} [1 + \mathrm{ln}(\dot{m}_{\rm 0})]  & \mbox{if } \dot{m}_{\rm 0} \geq 1 \\
		L_{\rm Edd} \dot{m}_{\rm 0} & \mbox{otherwise}
	\end{array}
\right.
\label{eq:Liso}
\end{equation}

\noindent Following \cite{Wiktorowicz_2017_Origin}, we defined the beaming factor, $b$:

\begin{equation}
b =
\left\{
	\begin{array}{ll}
		1  & \mbox{if } \dot{m}_{\rm 0} < 8.5 \\
		73/\dot{m}_{\rm 0}^{2} & \mbox{if } 8.5 \leq \dot{m}_{\rm 0} < 150\\
		3.2 \times 10^{-3} & \mbox{if } \dot{m}_{\rm 0} \geq 150 \\
	\end{array}
\right.
\label{eq:beaming_factor}
\end{equation}

\noindent which is related to the solid angle of the wind-cone subtending a
half-apex angle $\theta/2$ by $b = 1-\cos(\theta/2)$. A beaming factor of $b =
1$ corresponds to no beaming and a half opening angle $\theta / 2$ of
$90^{\circ}$, while the smallest, limiting value of $3.2 \times 10^{-3}$
corresponds to a half opening angle of $\sim 4.6^{\circ}$  (see
\citealt{Lasota_2016_beaming_saturation}). Under the assumption of an isotropic
volume distribution of sources, the beaming factor $b$ is equal to the probability of
observation, i.e. the probability that the beam enters our line-of-sight, which we account for
in our calculations. The $b \propto 1/\dot{m}_{\rm 0}^{2}$ relation was derived
by \citet{King_2009_Beaming} as an extension to the treatment of black-body
emission from the accretion disc around a black hole
\citep{King_2002_BH_black_bodies}, and supports the observation that several
bright ULXs appear to show an anti-correlation between their peak luminosity
and characteristic soft X-ray temperature \citep{Feng_2007_Lsoft_T4,
2009_Kajava_MNRAS.398.1450K}. 

With values calculated for $L_{\rm iso}$ and $b$, we then obtained the
{\it maximum} beamed luminosity $L_{\rm x}$
that would be observed for a given simulated system (noting the
aforementioned caveats) by dividing the intrinsic luminosity by its beaming
factor:

\begin{equation}
    L_{\rm x} \approx \frac{ L_{\rm iso} } { b }\\
	\label{eq:Lx}
\end{equation}

We note that, in the above, we have assumed the beamed luminosity corresponds
to the observed luminosity; this is an over-simplification (see the Discussion), as the true effect of
beaming on the spectrum (and therefore total X-ray luminosity) requires
consideration of the radial dependence of beaming (which can be substantially
different for regions around the spherisation radius compared to the inner-most
regions (Khan et al. in prep). We also note that the above formula assumes that the
flow is `classically' super-critical and therefore that the magnetic field of a
neutron star in a given ULX is typically weak enough such that the
magnetospheric radius is far smaller than the spherisation radius
(see
\citealt{Mushtukov_2017_PULXs_as_magnetars} for a discussion of the nature of
the flow when this condition is not met) and explicitly ignores emission
from the accretion column.

\subsection{Duty Cycle} \label{sec:duty_cycle}

The disc instability model (DIM; \citealt{2001_Lasota_NewAR..45..449L}), modified for irradiation, explains the outbursts of low-mass X-ray binaries (LMXBs) as being mediated by the
well-known thermal-viscous instability of an $\alpha$ disc (\citealt{Shakura_1973}, see also
\citealt{2020_Hameury_arXiv201000365H} for a recent extension to higher
accretion rates). The duty cycle of the resulting outbursts, $d$, is defined as
the fraction of total time spent in outburst. Observationally, the value for
$d$ is not particularly well constrained; extreme cases include that of
GRS~1915+105 with an outburst duration exceeding 20 years and a predicted
recurrence time of $\sim 10^{4}$ years, giving it an X-ray duty cycle of $\sim
0.1\%$ ($ d \sim 0.001$) \citep{2009_Deegan_MNRAS.400.1337D}. The Galactic
LMXB, GX 339-4 on the other hand has outbursts with a recurrence time of $\sim
450$ days; based on eight outbursts provided in
\cite{1998_Rubin_ApJ...492L..67R}, we estimate the duty cycle to be roughly
$\sim 30\%$ $(d \sim 0.3)$, whilst \textit{Chandra} observations of two ULXs in
NGC 5128 place an upper limit on their duty cycles of $d \sim 0.2$
\citep{2013_Burke_ApJ...775...21B}.


In order to accommodate the observational impact of duty cycles within our
parent population, we defined a sub-sample of systems undergoing nuclear
timescale mass transfer that are not wind fed, and had donor stars with an
effective temperature of $T_{\mathrm{eff}} < 7000K$ (to be below the
instability threshold, see \citealt{2001_Lasota_NewAR..45..449L}) and donor
star masses below $5 M_{\odot}$.  For simplicity, we make the assumption that
the outer regions of the disc have the same temperature as the companion star
calculated via $T_{\mathrm{eff}} = ({ L_{\mathrm{2}} / 4 \pi R_{\mathrm{2}}^2
\sigma)^{1/4} }$, where $\sigma$ is the Stefan-Boltzmann constant, and
$L_{\mathrm{2}}$, and $R_{\mathrm{2}}$ are the luminosity and radius of the
companion star respectively. For these systems likely to undergo outbursts
mediated by the DIM, we selected a single value for $d$ such that a system with, e.g.
$d = 0.2$, could potentially reach ULX luminosities for 20\% of its total
lifetime. At gas temperatures exceeding 7000~K, we assume that the disc is
constantly transporting angular momentum and does not display recurrent
outbursts i.e $d = 1.0$. With relevance to these latter sources, we note that
we have not included the propeller effect which may act to lower the duty cycle
for wind-fed or persistently accreting neutron star systems. We have also not
accounted for the duty cycles of Be-X-ray binaries which undergo recurrent
outbursts as a result of the neutron star's highly elliptical orbit and passage
through the decretion disc (e.g. \citealt{2011_Reig_Ap&SS.332....1R}).

\begin{figure}
    \includegraphics[scale=0.95]{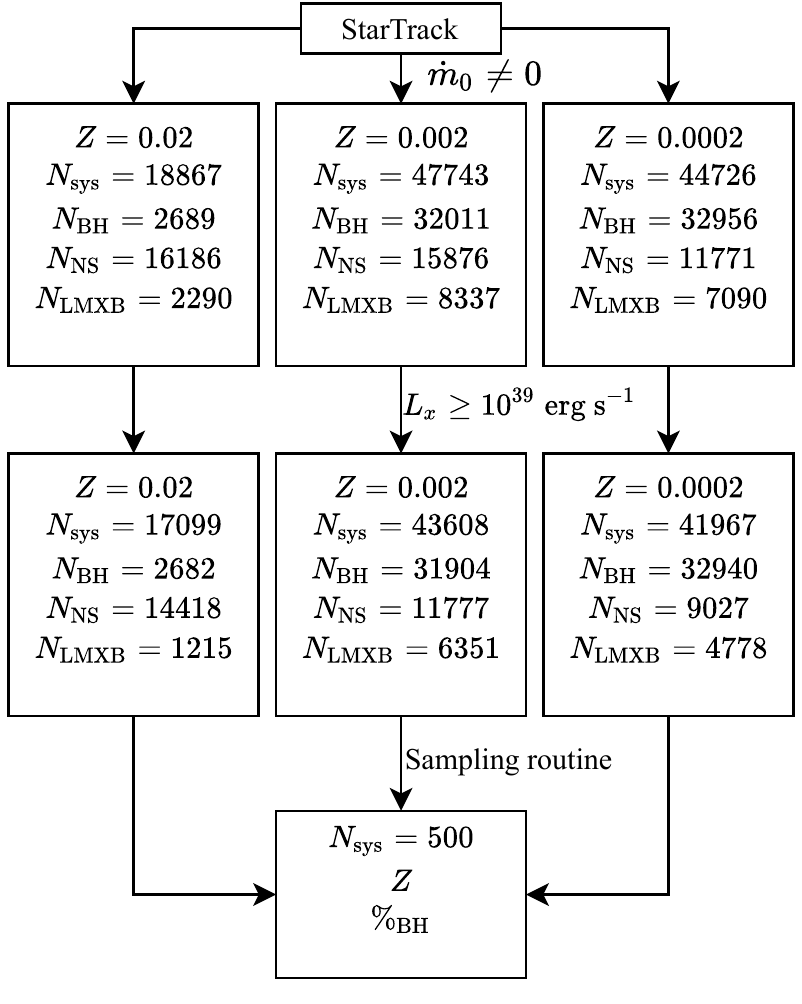}
    \caption{
        Flowchart showing the filtering process to obtain a representative sample
        population of ULXs. See section \ref{sec:samp_method} for more detail.
        }
    \label{fig:samp_diagram}
\end{figure}

\subsection{Obtaining a representative population of ULXs} \label{sec:samp_method}

The impact of precession on the observed population of ULXs depends on the
underlying population demographic, and thus we adopt the following method for
creating representative sample populations.  Figure \ref{fig:samp_diagram}
illustrates our method for generating samples of ULXs from the initial
populations created via \startrack.  The simulated results are grouped into
three metallicities: $Z = 1\%Z_\odot, \ 10\% Z_\odot, \ Z_\odot$ (where solar
metallicity, $Z_{\odot} = 0.02$) and we perform simulations for each group
separately as well as the combination of all three. As previously stated in
section \ref{sec:lum_b_fac}, we filtered to only include systems undergoing
active mass transfer. Next, we filtered out the 16,135 binary systems that were
undergoing mass transfer from a white dwarf or did not exceed $L_{x} =
1\times10^{39} \ \ergss$ at any point during their lifetime. We note that \cite{2014_Steele_ApJ...785..147S} argue that a globular cluster ULX in
NGC4472 \citep{2007Natur_Maccarone_.445..183M} displays findings consistent with a
white dwarf donor star (the primary being a black hole) and so it may be possible for such sources to exist,
however as their evolution is dominated by dynamic processes in the cluster, the inclusion of such ULXs requires specific prescriptions beyond the scope of this work. The filtering left us with 88,748 BH/NS systems that were undergoing either nuclear or
thermal timescale mass transfer and which serves as our parent ULX population. 
From here we split our sample into two groups depending on whether the compact
object is a NS or BH. This allowed us to later specify a desired black hole percentage
$\%_{\mathrm{BH}}$ within our population.

There are important differences in how long a given system may appear as a ULX, as
the duration of thermal timescale mass transfer is orders of magnitude shorter
than Roche lobe overflow on the nuclear timescale of the secondary (with repeat
outbursts mediated by the DIM). Therefore, should we sample {\it uniformly}
over all of the systems in our parent population, we would tend to
over-represent short-lived ULXs. To correct for this, we apply a sampling
prescription whereby the probability of selecting a given ULX is set by its
lifetime as a ULX, divided by the summed lifetime of all the other systems in
the parent population $P_{\mathrm{sample}} = t_{\mathrm{ULX, n}} /
\sum_{n=0}^{N_{\mathrm{ulx}}} t_{\mathrm{ULX, n}}$. This sampling procedure
explicitly assumes that there exists a constant star formation rate across
cosmic time.

In the following simulations, we have chosen to re-sample the total population
to produce smaller, volume-limited populations of $\Nsys = 500$ systems (for
each metallicity); this is a little larger than the currently \textit{observed}
number of ULXs by \textit{XMM-Newton} \citep{Earnshaw_ULX_cat} although through
repeat re-sampling (our Monte-Carlo procedure), the results can be scaled to
any desired population size.

\subsection{Simulating ULX light curves} \label{sec:lc_simulation}

In order to investigate the impact of precession on the observed population of
ULXs, we require a method for creating long term light curves for our
synthesised population. \texttt{ULXLC} \ is a numerical model developed by
\cite{Dauser_2017_ULXLC}, to describe the luminosity emerging from within an
optically thick wind-cone of half-opening angle $\theta / 2$, with an observer
inclination, $i$, and an outflow velocity (which also imparts some Doppler
boosting) of the outflow, $\beta_{\mathrm{wind}}$ (and which hereafter we fix to $0.3c$ to be
broadly consistent with the winds detected in ULXs and SS433:
\citealt{2014_winds,  2016_Walton_ApJ...826L..26W, Pinto_2016Natur.533...64P,
Pinto_2017AN....338..234P, 2018_Kosec_MNRAS.473.5680K,
2018_Kosec_MNRAS.479.3978K, SS433_2021}).

\texttt{ULXLC} also incorporates the effect of precession through a half
precessional angle $\Delta i$ and precession period $P$, which may be shifted
by a phase offset $\phi$. The model does not assume any physical mechanism for
{\it driving} the precession and so can be used without any additional {\it
a-priori} assumptions. In order to use \texttt{ULXLC}, we require input model
parameters; $\theta$ can be derived from the assumed relationship between accretion
rate, beaming and opening angle (Section \ref{sec:lum_b_fac}), whilst $\cos
i$ is uniformly distributed between zero and one, which ensures the random
distribution of ULX orientations in space. $\Delta i$ is an unknown,
although if we assume SS433 to be a reasonable indicator, values of
$20^{\circ}$ are not implausible \citep{1979_Fabian_MNRAS.187P..13F,
1979_Milgrom_A&A....78L...9M, 1979_Margon_BAAS...11..786M}.  Using
\texttt{ULXLC}, we simulated light curves for each ULX system, which took the
form of a single precession cycle and a time series of 5000 data points.

The light curves from \texttt{ULXLC} required normalising to produce physical
luminosity units. For any given combination of system parameters (i.e. our
simulated population of ULXs), we simulated a light curve at zero inclination
and set the maximum luminosity to be equal to the beamed luminosity given in
equation \ref{eq:Lx}. This allowed us to calculate a scaling constant which we
used to renormalise any light curve at arbitrary inclination and obtain a
luminosity in physical units. 

The light curves produced by \texttt{ULXLC} are periodic even though in reality precession may be quasi-periodic if dependent on accretion rate
(e.g. \citealt{Middleton_2019_Accretion_plane}). Although the period is not utilised
until we consider the regular observations taken by \erosita \ (section
\ref{sec:obs_pred_erass}), at the point of creating the light curves, we also
scale the period using formulae for Lense-Thirring precession
\citep{Middleton_2019_Accretion_plane} and an empirical-only relationship
\citep{Townsend_2020} (see equations \ref{eq:Period_wind} \& \ref{eq:P_sup}).

\subsubsection{Lense-Thirring Precession} \label{sec: P_wind_method}

Lense-Thirring precession is a relativistic correction to the precession of a
gyroscope near a large rotating mass. This effect has been shown to occur in
the weak-field limit around the Earth by the Gravity Probe B experiment
(\citealt{GravityProbeB}) and is believed to explain the type C quasi-periodic
oscillations in black hole binaries \citep{Stella_1998ApJ...492L..59S,
2012_Ingram_MNRAS.419.2369I, 2012_Ingram_MNRAS.427..934I,
2017_Ingram_MNRAS.464.2979I, lenseFlavours}.  The effect occurs when orbiting
fluid is displaced vertically from a rotating body's equatorial axis such that
frame-dragging then induces oscillations about the ecliptic and periapsis.
General relativistic magnetohydrodynamic (GRMHD) simulations of tilted
accretion discs around Kerr black holes (\citealt{Fragile_2007_tilted_disc})
display global solid body precession of the hot inner flow, and in ULXs it is
speculated that the same effect may lead to precession of the large
scale-height disc and wind cone (\citealt{Middleton_2018_Lense_Thirring,
Middleton_2019_Accretion_plane}).

Following \citet{Middleton_2019_Accretion_plane}, we calculated the precession
period of the wind-cone via equation \ref{eq:Period_wind}:

\begin{equation}
P_{\rm wind} = \frac{GM \pi}{3c^{3}a_{*}} r_{\rm out}^{3}
\left [   \frac{ 1 - \left ( \frac{r_{\rm in}}{r_{\rm out}} \right )^{3}}{\textup{ln} \left ( \frac{r _{\rm out} }{ r_{\rm in} } \right ) }   \right ]
	\label{eq:Period_wind}
\end{equation}

\noindent We make the simplifying assumption that $r_{\rm in} = r_{\rm isco}$
in units of the gravitational radius ($GM/c^{2}$), i.e. ignoring the role of
magnetic fields (although see \citealt{Vasilopoulos2019} and \citealt{Middleton_2019_Accretion_plane} for a
discussion). We assume that neutron stars are low spin ($a_{*} = 0.01, \ r_{\rm
isco} = 6 \ R_{g}$) as indicated by observations of pulsar ULXs (PULXs) to-date, with spin
periods of 1-10s of seconds \citep{King_Lasota_2020_PULX_iceberg_emerges}, and
that black holes may have very high spins ($a_{*} = 0.998, \ r_{\rm isco} =
1.25\ R_{g}$) as a consequence of the high accretion rates, the ability to
advect matter and angular momentum, and the lack of a propeller mechanism to
limit the spin-up. In the above, $r_{\rm out}$ is the outer photospheric radius
of the wind (the point at which radiation can free-stream) for which we assume
\citep{Poutanen_2007_ln}:

\begin{equation}
r_{\rm out} \approx \frac{3 \epsilon_{\rm wind}}{ \beta \zeta } \dot{m_{\rm 0}}^{3/2} r_{\rm isco}
	\label{eq:r_out}
\end{equation}

\noindent where -- for the purposes of determining this radius -- we have assumed $r_{\rm isco}
= 6 R_{g}$ for both NS and BHs (and an accretion efficiency of 0.08). Note that the discrepancy between assuming a high BH spin for the
precession period and a larger ISCO radius here does not have a substantial
effect on the location of the photosphere for large accretion rates (see
\citealt{Middleton_2019_Accretion_plane} for details). In the above, $\beta$ is
the ratio of asymptotic wind velocity relative to the Keplerian velocity at
$r_{\rm sph}$, and, for simplicity, we set this to $1.4$.  $\epsilon_{\rm wind}
= L_{\rm wind}/L_{\rm tot}$ is the fraction of dissipated energy used to launch
the wind, which we set to $\epsilon_{\rm wind} = 0.25$ (\citealt{Jiang_2014},
see also \citealt{Pinto_2016Natur.533...64P} for a higher inferred value from
observation). Finally, $\zeta$ is the cotangent of the opening angle of the wind cone
which we assume is equal to:

\begin{equation}
\zeta = \mathrm{tan}\left [ \frac{\pi}{2} - \mathrm{acos} \left (  1 - b \right )   \right ]
\label{eq:zeta}
\end{equation}

\noindent We assume a lower-limit of $\zeta = 2$ based on radiative
magnetohydrodynamic simulations at moderate super-Eddington rates
(\citealt{Sadowski_2014}, and noting that in reality, $\zeta$ -- and therefore $b$ -- likely increases
in a more complicated fashion with $\dot{m_{\rm 0}}$: \citealt{Jiang_2019}).

\subsubsection{Empirical Precession} 

In addition to the above physical precession mechanism, we also utilise the
result of \cite{Townsend_2020}, where the mechanism for precession is unknown
but the super-orbital (P$_{\rm sup}$) and orbital periods (P$_{\rm orb}$)
are inferred to be related by:

\begin{equation}
    P_{\rm sup} = 22.9 \pm 0.1 P_{\rm orb}
	\label{eq:P_sup}
\end{equation}

\noindent where P$_{\rm orb}$ is given by:

\begin{equation}
    P_{\rm orb}  = 2 \pi \sqrt{ \frac{a^3}{G(M_{\rm c}+M)} }     
	\label{eq:P_orb}
\end{equation}

\noindent where $M_{\rm c}$ is the mass of the companion star and $a$ is the
semi-major axis of the binary system.

\subsection{Effects of precession on the observed population of ULXs} \label{sec:MC_method}

As we mention above, in order to explore the impact of various key parameters
on our observations of ULXs, we re-sample the parent population 10,000 times,
each time producing a smaller sample of 500 ULXs.  For each ULX in our smaller
sample, we provided the following parameters to \textsc{ulxlc}: $\theta / 2$,
$i$, $\Delta i$, $L_{\rm x}$, $P$, $\phi$, $\beta_{\mathrm{wind}} = 0.3$. $\Delta i$ and $\phi$
are sampled from uniform distributions with the following ranges: $0^\circ \le
\Delta i \le \dimax$, $0 \le  \phi \le 1$. 
$\cos i$ is uniformly distributed between zero and one, $\theta / 2$, whilst $L_{x}$ and $P$
are calculated quantities of the particular system.  We explore the impact of $\Delta
i_{\mathrm{max}}$ = $45^\circ$ and $20^{\circ}$, as we do not have strong
constraints on the precessional angle (other than for SS433). We then proceeded to
classify each light curve in our sample, created using \textsc{ulxlc}, into one
of three categories:

\begin{itemize}
  \item \makebox[1.5cm]{\textit{Alive}:\hfill} Persistently above $1\times10^{39} \ \ergss$
  \item \makebox[1.5cm]{\textit{Transient}:\hfill} Systems that crossed $1\times10^{39} \ \ergss$
  \item \makebox[1.5cm]{\textit{Hidden}:\hfill} Persistently below $1\times10^{39} \ \ergss$
\end{itemize}

\noindent We note that this act of classifying sources is independent of the
precession period, with systems merely being defined based on the above,
regardless of the timescales involved.  The numbers of systems in each
classification were recorded and saved such that the total number of systems
($\Nsys$ = 500) = the number alive ($\Nalive$) + the number of transients
($\Ntransient$) + the number of hidden systems ($\Ndead$). Light curves that
were classified as \textit{transient} were subjected to further analysis (see
Section \ref{sec:erass_sampling_routine}), whilst ULX systems with half opening
angles of $\theta/2 > 45^{\circ}$ (set by the accretion rate -- see equation
\ref{eq:beaming_factor}) were considered to be sources that do not display
precession (see \citealt{Dauser_2017_ULXLC}) and thus were classified as being
alive without the need to simulate light curves.

\subsection{Simulations of the X-ray Luminosity Function} \label{sec:XLF_method}

X-ray luminosity functions (XLFs) -- both in their differential and cumulative
forms -- have been commonly extracted from survey data in order to study
population demographics \citep{1989_Fabbiano_ARA&A..27...87F, 2003_Grimm_MNRAS.339..793G, 2016_wang_ApJ...829...20W}. XLFs can provide insights
into the star-formation history \citep{2013_Fragos_ApJ...764...41F,
Fragos_2013_reionization} and impose constraints on theoretical models of
binary evolution. It is important to consider how the combination of precession
and geometrical beaming can together affect the overall shape of observed XLFs.
To this end we explored how a synthetic XLF would appear under six different
scenarios that could describe a given source's luminosity:

\begin{itemize}
    \item $L_{\mathrm{iso}}$: the isotropic luminosity obtained in the absence of beaming (eq. \ref{eq:Liso})
    \item $L_{\mathrm{x}}$: the above including geometrical beaming (eq. \ref{eq:Lx})
    \item $L_{\mathrm{x,b}}$: the above including the probability of observation set by the beaming factor (i.e. assuming obscuration by the wind)
    \item $L_{\mathrm{x,b*d}}$: the above including the additional effect of the LMXB duty cycle on the observation probability
    \item $L_{\mathrm{x,prec}}$: the luminosity obtained via the generation and uniform sampling of light curves produced by \ulxlc \ (sec. \ref{sec:lc_simulation})
    \item $L_{\mathrm{prec, vis}}$: the above including the additional effect of the LMXB duty cycle.
\end{itemize}

As described in section \ref{sec:lum_b_fac}, we re-sampled $\Nsys = 500$
binaries from across all metallicites from our \textbf{full} parent population
weighted by their lifetimes in the active mass transfer phase (top row in
Figure \ref{fig:samp_diagram}), while specifying a black hole percentage of
$\%_{\mathrm{BH}} = 0, \ 50\%$ \& $100\%$. For the construction of our XLFs, we separate the
BHs, NSs, and LMXBs, as well as those defined as alive or transient;  this is
useful for illustrating the relative contributions of each component to the XLF.

LMXB sources were set to have a duty cycle of $d=0.2$ so that there was a 20\%
chance of them being observed at their luminosity given by $L_{\mathrm{x}}$,
and an 80\% chance of them not being observed at all (i.e. $L_{\mathrm{x}}=0$).
Thermal timescale or wind-fed systems in our population were set to have a duty
cycle of $d=1.0$. For each system (with $\theta/2 < 45^{\circ}$) a light curve
was then generated using {\sc ULXLC} assuming a precessional
angle uniformly sampled between $0$ and $\dimax=45^{\circ}$; we then randomly
sampled the system's light curve to obtain its new luminosity. Re-sampling the
parent population (as described in section \ref{sec:samp_method}),
then allows us to obtain 1$\sigma$ errors on each luminosity bin for any given
XLF.

\subsection{Simulating \erass's view of the ULX population}\label{sec:eRASS_trans}

The \erosita \ X-ray telescope was launched in July 2019 and has already begun
its all-sky survey, \erass, which takes snapshots of the entire sky in the $0.2
- 10 \ \mathrm{keV}$ band, repeating every six months for a period of four
years \citep{eRositaScience}. Using the generated light curves for our artificial population of ULXs,
we can obtain predictions for what \erosita \ might observe given an underlying
population demographic, and explore how our constraints might improve over the course of
the \erass \ four year survey.

Whilst the previous sections had no requirement to use the periods given by
equations \ref{eq:Period_wind} \& \ref{eq:P_sup}, given \textit{eRASS's}
regular observations it is important to factor in the deterministic nature of
such a periodic (or, in reality quasi-periodic --
\citealt{Middleton_2019_Accretion_plane}) modulation of the luminosity. 

We note that our simulations make the assumption that all the systems in our
parent population have an equal probability of observation regardless of their
spatial distribution, luminosity or spectra.  In reality, there will be a
natural bias towards detecting the brighter sources, which is further compounded by
the anisotropic sensitivity of \erass \ (with greater effective exposure
occurring near the ecliptic poles and deeper coverage between 0.2
- 2.3 \ \rm{keV},  \citealt{2020_Predehl_arXiv201003477P}).
To obtain a more realistic picture requires the distribution of simulated
binaries amongst galaxies out to a few 10s of Mpc, some estimate for the true
number per galaxy type (and per unit star formation), their spectra and the
convolution of the exposure time and detector response.  Whilst this is beyond
the scope of this work, we discuss the impact of resulting bias in the
Discussion section.

\subsubsection{\erass \ Sampling Routine} \label{sec:erass_sampling_routine}
The light curves that were created in Section \ref{sec:MC_method}, were scaled
to have a period of both $P_{\rm wind}$ and $P_{\rm sup}$ and their luminosity
was then sampled in intervals of six months to match the observing cadence of
\erass.  At each \erass \ cycle $(c)$, we keep track of the following: 

\begin{itemize}
    \item Sources above $1\times10^{39} \ \ergss$, $\Nulx$
    \item Sources below $1\times10^{39} \ \ergss$, $\Nnotulx$
    \item Newly detected ULXs, $\Nnew$
    \item Previously detected ULXs that fell below the ULX threshold, $\Ndip$
    \item The change in the number of ULXs $\Ndeltaulx = \Nnew - \Ndip$
    \item The number of transient sources, $\Ntransients = \Nnew + \Ndip$ (for $c > 1$)
    \item The number of alive sources, $\Nalives$
\end{itemize}

The above quantities are naturally cycle-specific and the cumulative equivalents for these quantities may be obtained by summing over all \erass \ cycles, e.g. we define the
cumulative number of observed sources by $\Nobs = \csum \Nnew$.
We also note that the quantity $\Nalives$
includes the $\Nalive$ systems classified as \textit{alive}, with opening angles $\theta/2>45^{\circ}$, and for which light curves were not simulated (see section 
section \ref{sec:MC_method}).

Over the first 6 months of \erass \ (cycle 1), we make the assumption that the
survey will not detect any transient sources due to precession, as the exposure
time is very short relative to the typical precession timescale. At the conclusion of
cycle 1 we therefore have a starting value for the total number of observed
ULXs which will subsequently increase as the survey continues.

We performed 10,000 sets of Monte Carlo simulations for a given combination of
input parameters which covered $Z$ (0.02, 0.002, 0.0002 and the combination of
all three), $\%_{\rm{BH}}$ (0 $\rightarrow$ 100 \% in steps of 25\%), $\Delta
i_{\rm{max}}$ (20$^{\circ}$ and 45$^{\circ}$), P ($P_{\rm{wind}}$ and
$P_{\rm{sup}}$) and d (0.2 and 1.0).  At each \erass \ cycle we recorded
quantities which may be compared to actual \erass \  measurements, such as the
number of sources detected in a given \erass \  cycle $\Nulx$.  The repeat
simulations allowed the construction of distributions from which we extracted
the key statistics related to the various quantities in each cycle as a
function of our physical parameters (notably $\%_{\rm BH}$). 

An example set of results from a
single Monte-Carlo iteration is shown in Table \ref{tab:eRASS}.

\begin{table}[h]
\centering
\resizebox{\columnwidth}{!}{\begin{tabular}{lllllll}
\hline
 c &  $\Nnew$ &  $\Ndip$ &  $\Ndeltaulx$ & $\Ntransients$  &  $\Nalivecum$ &  $\Nobs$ \\
\hline
1 & 303 & 0 & +303 &  0  & 303 & 303 \\
2 &  24 & 12 & +12 & 36  & 291 & 327 \\
3 &   8 & 4 &  +4 &  12  & 287 & 335 \\
4 &   3 & 0 &  +3 &  3  & 287 & 338 \\
5 &   2 & 2 &  0  &  4  & 285 & 340 \\
6 &   1 & 0 &  +1 &  1  & 285 & 341 \\
7 &   1 & 2 &  -1 &  3  & 283 & 342 \\
8 &   2 & 0 &  +2 &  2  & 283 & 344 \\
\hline
\end{tabular}}
\caption{An example of a single \erass \ Monte-Carlo iteration showing a subset of observed quantities, created from an underlying
population of $\Nsys = 500$ ULXs with $\%_{\rm BH} = 50$, $\Delta i_{max} = 20^{\circ}$,
$Z = 0.02$ and $d = 1.0$. The numbers illustrate the evolution of the observed
population as seen by \erass \ over its 8 cycles.}\label{tab:eRASS}
\end{table}

\section{Results}

\subsection{The Impact of Precession on the XLF}

\begin{figure*}
    \includegraphics[scale=1.1]{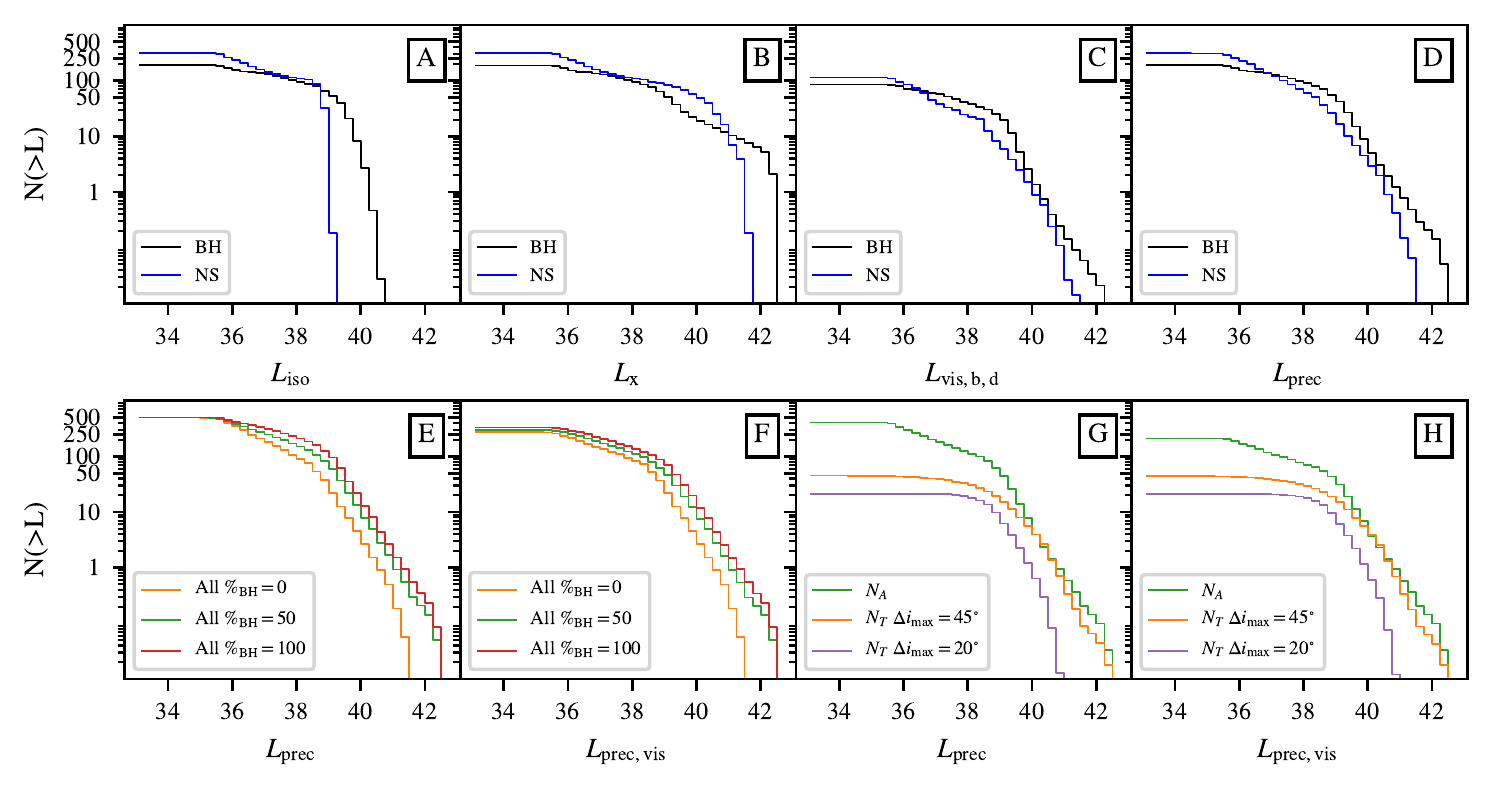}
    \caption{Cumulative XLFs shown for a variety of luminosity prescriptions.
    \textbf{A:} the XLF for BHs and NSs when assumed to emit isotropically ($L_{\rm iso}$).
    \textbf{B:} the same for beamed emission ($L_{\rm x}$).
    \textbf{C:} The same but factoring in the observation probability from beaming and LMXB duty cycle (but without precession: $L_{\mathrm{vis,b,d}}$).
    \textbf{D:} the XLF after incorporating the probability obtained from precession
    $L_{\mathrm{prec}}$. (see section \ref{sec:modelling_xlf}). 
    \textbf{E:} $L_{\mathrm{prec}}$ for different $\%_{\mathrm{BH}}$.
    \textbf{F:} The XLF incorporating precession and the addition of a LMXB duty cycle
    ($L_{\mathrm{prec,vis}}$).
    \textbf{G:} The XLF incorporating precession split into the classifications
    described in section \ref{sec:MC_method}.
    \textbf{H:} The XLF incorporating precession and duty cycle split into the classifications
    described in section \ref{sec:MC_method}.
    ($L_{\mathrm{prec,vis}}$).
    The first four and last two panels assume $\%_{\mathrm{BH}}=50$, while the estimates for the impact of precession assume a maximum
    precessional angle of $\dimax = 45^{\circ}$ unless otherwise stated.
    Where we have included its effect, the LMXB duty cycle was set to $d=0.2$.
    The XLF obtained via sampling from the ULX catalogue of
    \protect\cite{Earnshaw_ULX_cat} is shown in cyan.
    The method for the creation of this plot is detailed in Section \ref{sec:XLF_method}.
    } 
    
\label{fig:XLF_BH_NS}
\end{figure*}

Following the method detailed in Section \ref{sec:XLF_method}, Figure
\ref{fig:XLF_BH_NS} shows several of our synthetic cumulative XLFs, created
using the method described in section \ref{sec:samp_method}. We now use the
total lifetime of the source during active mass transfer as opposed to the
lifetime of the source during only the ULX phase, so that the sampling probability
is given by $P_{\mathrm{sample}} = t_{\mathrm{mt, n}} /
\sum_{n=0}^{N} t_{\mathrm{mt, n}}$, where $t_{\mathrm{mt, n}}$ is the
amount of time spent undergoing active mass transfer for the nth source, and $N$
is the number of sources in the population.

Panel \textbf{A} shows
an XLF assuming $\%_{\rm{BH}}$ = 50\% and isotropic emission in the absence of
any geometrical beaming, with no neutron stars exceeding the $1\times10^{39} \
\ergss$ luminosity threshold (as the beaming only begins at a NS luminosity
around 6 $\times$ 10$^{38} \ \ergss$), and a few BHs reaching up to $\sim 10^{40} \
\ergss$. 

Panel \textbf{B} (also for $\%_{\rm{BH}}$ = 50\%) shows the same XLF as in Panel \textbf{A}
after incorporating beaming but does not account for the observation
probability (i.e. it assumes every detected source is observed directly down
the wind cone). We see that, between $10^{38}$ and $\sim 10^{41} \ \ergss$, NSs
appears to dominate, while at the highest luminosities, $>10^{41} \ \ergss$, BH
accretors dominate.  

Panel \textbf{C} includes the observation probability
provided by the beaming factor $b$, and the LMXB duty cycle $d$; we see that
the brightest sources above $\sim 10^{40} \ \ergss$ are suppressed and are no
longer visible.  

Panel \textbf{D} (also for $\%_{\rm{BH}}$ = 50\%) includes the
combined effects of geometrical beaming, precession (via ULXLC) and a LMXB duty
cycle of $d=0.2$; both NS and BH systems are observed in similar numbers across
the full range of luminosities, with none detected above $\sim 10^{41} \
\ergss$.  

Panels \textbf{E} and \textbf{F} are created following the same
process as panel \textbf{D} and show the XLF with and without the addition of
the LMXB duty cycle respectively. Here we have combined the NS and BH
populations into a single observed population and varied $\%_{\rm BH}$.  We
observe that the general shape of the XLF is not strongly affected by the
underlying $\%_{\rm BH}$, however, for a higher proportion of BHs in the
underlying population, there are a larger number of detected systems across all
luminosities (with systems still being detected at a few $\times 10^{41} \
\ergss$).

Panels \textbf{G} and \textbf{H} show the same as panels \textbf{E}
and \textbf{F} except we have now split the XLF into the classifications of
alive and transient, with results shown for different maximum precessional
angles as described in section \ref{sec:MC_method}.

\begin{figure}
\centering
\includegraphics{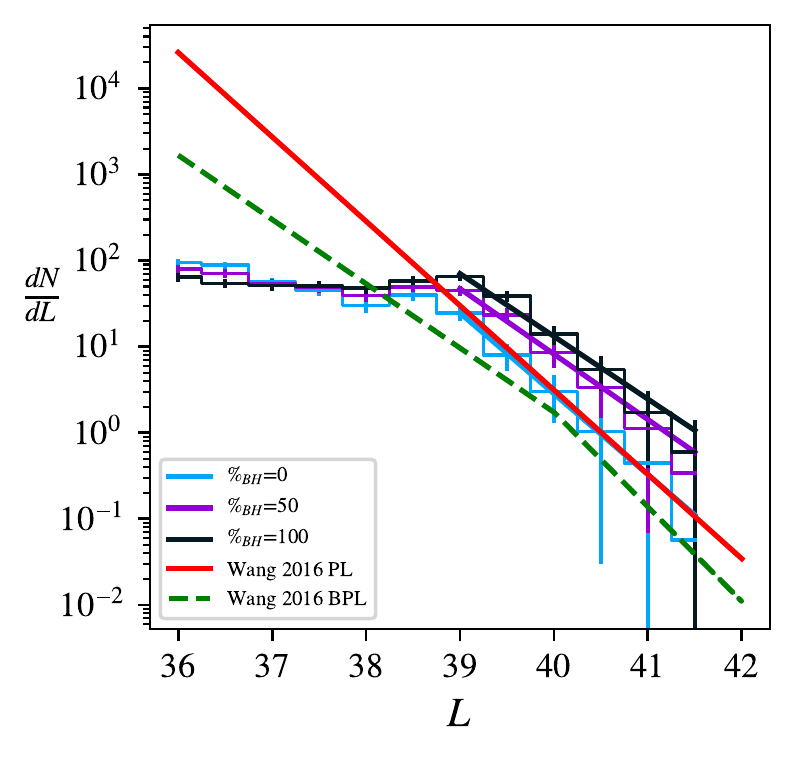}
\caption{
Several realisations of our synthetic, differential XLFs (showing $L_{\mathrm{prec}}$), for different black hole ratios: $\%_{\mathrm{BH}} = 0$ (cyan), 50 (magneta)
and 100 (black) and a maximum precessional angle of $\dimax=45^{\circ}$. Best fit models (both broken and unbroken power-law) from  \citep{2016_wang_ApJ...829...20W}
are overlaid for comparison.
}
\label{fig:xlf_comparison}
\end{figure}

\subsubsection{Modelling the XLF} \label{sec:modelling_xlf}
The differential forms of observed XLFs $(dN/dL)$ are often
fitted with power-laws, or variants such as broken or exponential cutoff power-laws
(see \citealt{2003_Grimm_MNRAS.339..793G, Swartz_2011_Complete_sample_of_ULXs, 2012_Mineo_MNRAS.419.2095M, 2016_wang_ApJ...829...20W, 2018_Wolter_ApJ...863...43W, 2020_Kovlakas_MNRAS.498.4790K}). In Figure \ref{fig:xlf_comparison} we plot a subset of the differential forms of our synthetic XLFs versus the best fit models from \cite{2016_wang_ApJ...829...20W} who used \textit{Chandra} observations of 343 galaxies
(totalling 4970 sources, 218 of which are ULXs) to create differential XLFs. Whilst there appears to be agreement at higher luminosities $(L>10^{39})$,  at lower luminosities our XLFs appear to flatten off which is
inconsistent with the models based on observation; this is likely due to the excluded systems from our sampling which emit at lower luminosities (e.g. white dwarf accretors). In order to make a fair, quantitative comparison to reported slopes in the literature, we therefore model only the high luminosity tail
($>$ 10$^{39} \ \ergss$) of the differential form of our synthetic XLFs. 

Our differential luminosity functions obtained via $L_{\mathrm{\rm prec}}$ and
$L_{\mathrm{\rm prec,vis}}$ are fitted using a power-law of the form $dN/dL = AL^{-\alpha}$ via a method of maximum likelihood (for limitations on this
method see \citealt{2007_Clauset_arXiv0706.1062C}) where the errors on each bin are the standard deviation (rather than standard error which are considerably less representative in this case). Table \ref{tab:xlf_fit}
reports the corresponding best fit parameters and their 1$\sigma$ errors. 


\begin{table}[h]
\begin{tabular}{lrrcc}
\hline
L & $\dimax$ & $\%_{\mathrm{BH}}$ & $A$ & $\alpha$ \\
\hline
$L_{\mathrm{prec}}$ & 45 & 0 & 24.60 $\pm$ 0.77 & 0.94 $\pm$ 0.03 \\
$L_{\mathrm{prec}}$ & 45 & 25 & 36.18 $\pm$ 1.71 & 0.81 $\pm$ 0.03 \\
$L_{\mathrm{prec}}$ & 45 & 50 & 47.16 $\pm$ 3.00 & 0.76 $\pm$ 0.04 \\
$L_{\mathrm{prec}}$ & 45 & 75 & 59.23 $\pm$ 4.56 & 0.74 $\pm$ 0.05 \\
$L_{\mathrm{prec}}$ & 45 & 100 & 70.40 $\pm$ 6.13 & 0.73 $\pm$ 0.06 \\
$L_{\mathrm{prec}}$ & 20 & 0 & 23.95 $\pm$ 0.45 & 1.06 $\pm$ 0.02 \\
$L_{\mathrm{prec}}$ & 20 & 25 & 35.61 $\pm$ 1.79 & 0.88 $\pm$ 0.04 \\
$L_{\mathrm{prec}}$ & 20 & 50 & 47.36 $\pm$ 3.38 & 0.83 $\pm$ 0.05 \\
$L_{\mathrm{prec}}$ & 20 & 75 & 59.86 $\pm$ 5.53 & 0.79 $\pm$ 0.06 \\
$L_{\mathrm{prec}}$ & 20 & 100 & 71.20 $\pm$ 7.33 & 0.78 $\pm$ 0.07 \\
$L_{\mathrm{prec,vis}}$ & 45 & 0 & 24.24 $\pm$ 0.76 & 0.93 $\pm$ 0.03 \\
$L_{\mathrm{prec,vis}}$ & 45 & 25 & 29.28 $\pm$ 1.42 & 0.77 $\pm$ 0.03 \\
$L_{\mathrm{prec,vis}}$ & 45 & 50 & 34.06 $\pm$ 2.34 & 0.70 $\pm$ 0.04 \\
$L_{\mathrm{prec,vis}}$ & 45 & 75 & 39.61 $\pm$ 3.43 & 0.66 $\pm$ 0.05 \\
$L_{\mathrm{prec,vis}}$ & 45 & 100 & 44.44 $\pm$ 4.64 & 0.64 $\pm$ 0.06 \\
$L_{\mathrm{prec,vis}}$ & 20 & 0 & 23.60 $\pm$ 0.44 & 1.06 $\pm$ 0.02 \\
$L_{\mathrm{prec,vis}}$ & 20 & 25 & 28.63 $\pm$ 1.26 & 0.84 $\pm$ 0.03 \\
$L_{\mathrm{prec,vis}}$ & 20 & 50 & 34.19 $\pm$ 2.58 & 0.77 $\pm$ 0.05 \\
$L_{\mathrm{prec,vis}}$ & 20 & 75 & 39.98 $\pm$ 3.95 & 0.72 $\pm$ 0.06 \\
$L_{\mathrm{prec,vis}}$ & 20 & 100 & 44.96 $\pm$ 5.39 & 0.69 $\pm$ 0.07 \\
\hline
\end{tabular}
\caption{Best fit parameters and 1$\sigma$ errors from modelling our synthetic, differential XLFs above $10^{39}$ erg/s using a power
law of the form $AL^{-\alpha}$.}
\label{tab:xlf_fit}
\end{table}

As can be seen from Table \ref{tab:xlf_fit}, we observe a slight flattening of
the XLF slope with increasing $\%_{\mathrm{BH}}$ with slightly steeper slopes found for
$\dimax = 20^{\circ}$ when compared to $\dimax = 45^{\circ}$. The effect of the
duty cycle is to lower the maximum height reached by the XLF (i.e. the total number
of sources, see bottom row in Figure \ref{fig:XLF_BH_NS}) which flattens the slope,
especially when the population is BH dominated and thus extends to higher luminosities.

The existing literature contains a great deal of variation in the normalisation
when fitting functional forms to XLFs.  However the slopes of our synthetic differential XLFs
($\alpha$) are found to be somewhat flatter than those found in \citep{2003_Grimm_MNRAS.339..793G}
(created from HMXBs in five different galaxies), with an observed slope of $\alpha = 1.61 \pm 0.12$, in \citep{Swartz_2011_Complete_sample_of_ULXs} (using observations of 127 nearby galaxies) with an observed slope of $\alpha = 1.4 \pm 0.2$ above $10^{39} \ergss$, and in \citep{2016_wang_ApJ...829...20W} who applied a broken power-law (with a break at
$L_{b} = 2.5\times 10^{38} \ \ergss$), finding the slope above the break to be
$\alpha_{2} = 1.1 \pm 0.02$. We discuss the impact of observational bias as
the likely reason for this difference in the Discussion section.


\subsection{Dependence of Light Curve Classifications on Model Parameters} \label{sec:lc_dep_on_params}

Following from our simulations and the placing of sources into the three
categories described in Section \ref{sec:MC_method}, we now describe how the
underlying nature of the population might affect our observations of ULXs.

Figure \ref{fig:corner_adt_bh} shows the distributions for the number
of our three light curve classifications, as well as the percentage of transient to
total observable $\Ntransient / (\Nalive + \Ntransient)$ systems. The results
are presented as a corner plot \citep{2021_Dan_Foreman-Mackey} over a subset
grid of simulation parameters (see section \ref{sec:MC_method}).
The two distinct regions of parameter space in Figure
\ref{fig:corner_adt_bh} denoted by dotted and solid contours arise from the two different
maximum precessional angles, $\dimax$ $20^{\circ}$ (dotted) \& $45^{\circ}$ (solid).
There is considerable overlap in the number of alive and hidden systems from populations
drawn when using a maximum precessional angle of $20^{\circ}$ when compared to $45^{\circ}$.
However the error contours describing the number of transient sources, $\Ntransient$, overlap less, with smaller maximum precessional angles (up to $20^{\circ}$) resulting in
fewer transient sources by around a factor two when compared to the larger maximum precessional
angle (up to $45^{\circ}$).
The darker regions in Figure \ref{fig:corner_adt_bh} correspond to populations drawn
with higher fractional abundances of BHs, while blue-er regions correspond to populations
with a higher abundance of NSs. We observe that the number of hidden and transient sources,
$N_{\rm Hid}$ and $N_{\rm T}$, are negatively correlated with increasing black
hole percentage, while the number of alive systems increases with increasing
$\%_{\mathrm{BH}}$, this trend is observed across all of our simulated metallicities,
maximum precessional angles and simulated duty cycles.
The latter observation follows naturally from the expectation that NS ULXs are beamed (under our
assumptions which do not factor in the emission from the column nor the effects of strong
dipole fields) and, combined with precession, are more likely to be observed as {\it transient}
or {\it hidden}. We note that, in the absence of a LMXB duty cycle
(i.e. $d=1.0$), the correlation between the number of transient sources and black hole percentage
is markedly less pronounced, this is due to our prescription for LMXB systems
(section \ref{sec:duty_cycle}) resulting in a higher number of black hole systems displaying
outburst duty cycles when compared to NS systems.

In terms of the most extreme scenarios, from Figure
\ref{fig:corner_adt_bh} we can see that for a population composed entirely of neutron stars,
around $\sim 40 - 50\%$ of the observable sources are defined as being transient for
$\dimax=20^{\circ}$, or $\sim 60 - 75\%$ for $\dimax=45^{\circ}$. Conversely, for an
underlying population composed entirely of black holes, the proportion of sources being
defined as transient is $\sim 10-25\%$ for $\dimax=20^{\circ}$ or $\sim30-45\%$ for
$\dimax=45^{\circ}$.

\begin{figure}
\centering
\includegraphics{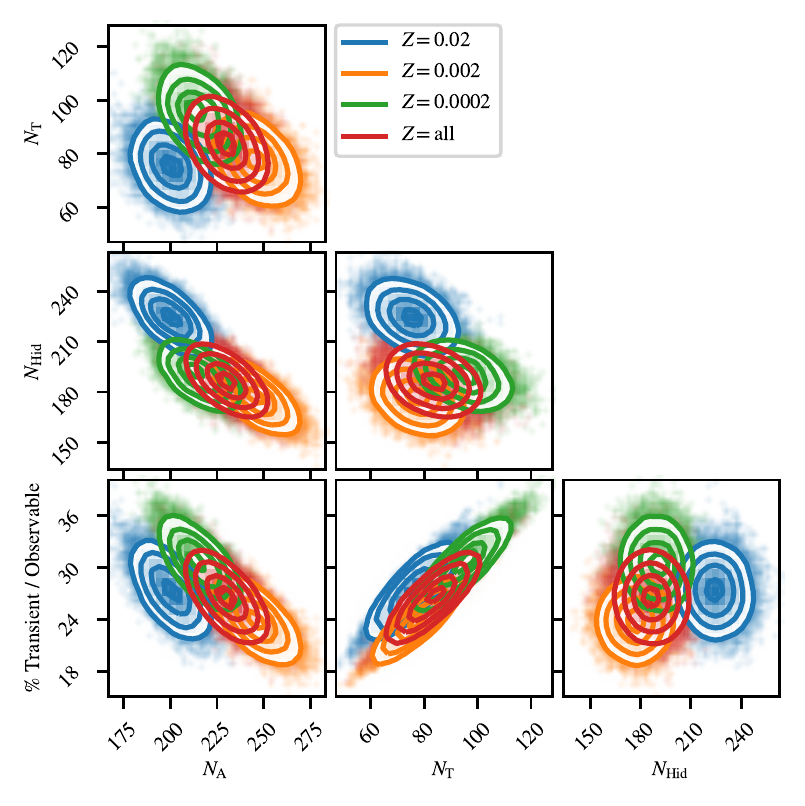}
\caption{
2D 1$\sigma$ error contours for the distributions of light curve classifications
for different parent population metallicities, $Z=0.02$ (blue), $Z=0.002$ (orange),
$Z=0.0002$ (green)
and the combination of all three (red) for fixed parameters:
$\dimax = 20^{\circ}$, $d = 0.2$, $\%_{\mathrm{BH}}= 50$.
}
\label{fig:adt_corner_diff_z}
\end{figure}

The effect of changing the parent population's metallicity, $Z$, which is shown
for a fixed set of model parameters in Figure \ref{fig:adt_corner_diff_z}, does have
an impact on the absolute numbers of each classification, however, the general trends
previously described hold true for all metallicities and their combination.


\subsection{Comparison to observations} \label{sec:comp_to_obs}
From Figure \ref{fig:corner_adt_bh} we deduce that the black hole percentage in the underlying population, and maximum precessional angle, substantially affects the percentage of transient to observed sources.
This implies that, with constraints on the maximum precessional angle and suitable
coverage (both in terms of area observed and cadence), it may be possible to
constrain the ratio of BHs to NSs in the underlying population simply by
determining the ratio of transient to alive systems (under the assumption that
the variability is driven by precession and the beaming is highly sensitive to accretion rate -- see the Discussion).

In the following sections we discuss initial constraints from \textit{XMM-Newton}
and then discuss implications for \erosita \ and \erass.

\subsubsection{Constraints from \textit{XMM-Newton}}

\begin{figure}
    \centering
    \includegraphics{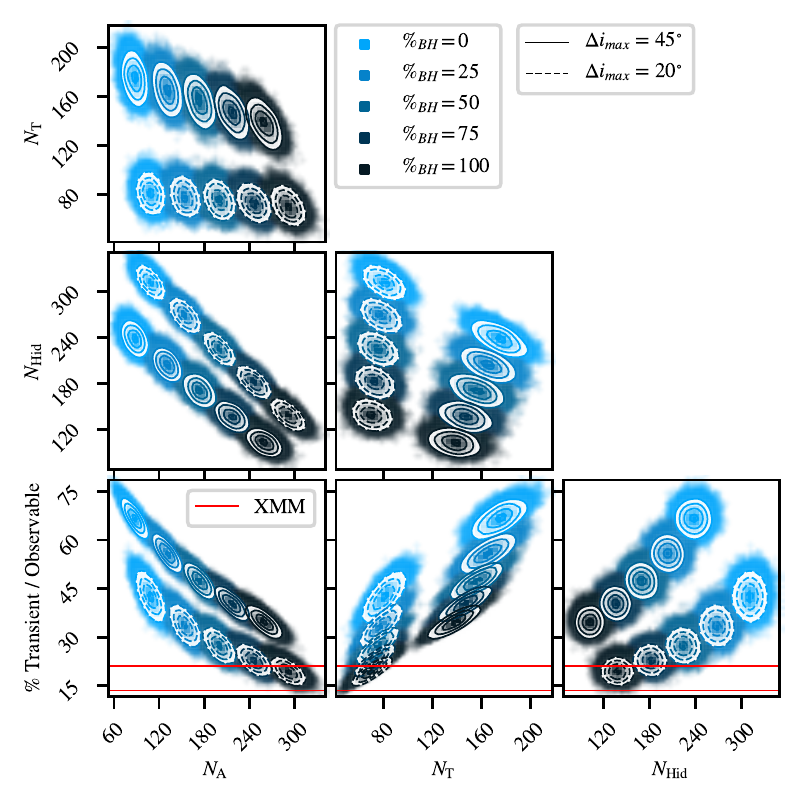}
    \caption{
    Shown are 2D 1$\sigma$ error contours for the distributions of the number of alive,
    transient and hidden systems ($\Nalive$, $\Ntransient$, $\Ndead$), as
    well as the percentage of transient to observed systems,
    ($\Ntransient / (\Nalive + \Ntransient))$ and how these vary with the black hole percentage
    ($\%_{\mathrm{BH}}$) of the underlying population.
    This particular simulation used the following fixed parameters:
    $\Nsys=500$, $Z=0.02$, $\dimax = 20^{\circ}$ (dashed contours)
    $45^{\circ}$ (solid contours) and $d=0.2$. $\%_{\mathrm{BH}}$ was varied between 0, 25, 50, 75 \& 100\%, where 
    a higher abundance of
    BH systems is shown on the figure as darker colours and blue-er colours correspond to higher abundances of NS systems.
    The red lines denote the $1\sigma$ confidence bounds for
    the percentage of transient systems to observable systems obtained from the most recent
    \textit{XMM-Newton} ULX catalogue \citep{Earnshaw_ULX_cat}
    }
    \label{fig:corner_adt_bh}
\end{figure}

To obtain some initial observational constraints on the number of alive and
transient ULXs, we used the catalogue of 1314 X-ray sources compiled by
\cite{Earnshaw_ULX_cat}, created from the 3XMM-DR4 data release of the
\textit{XMM-Newton} Serendipitous Source Catalogue
\citep{2016_Rosen_A&A...590A...1R}.  The catalogue identifies 384 candidate
ULXs, 81 of which were observed more than once. Each entry within the catalogue
includes a full band (0.2 - 12~keV) apparent (absorbed) luminosity and
associated $1\sigma$ errors.  From the 81 ULXs with multiple observations, we
sampled the luminosity (i.e. using their associated errors) 100,000 times, and
separated these systems into \textit{alive} or \textit{transient} based on our
previous definitions, and calculated associated $1\sigma$ error intervals on
the respective distributions. We find that $81 \pm 12\%$ of the systems may be
classified as \textit{alive}, while $19 \pm 3.8\%$ may be classified as
\textit{transient}; by comparison to our simulated results, we can thereby
obtain a crude estimate of the underlying, intrinsic properties of the observed
population. The region denoted by the red lines on Figure \ref{fig:corner_adt_bh}
indicates the 1$\sigma$ interval for the percentage of transient to observed sources
obtained from \cite{Earnshaw_ULX_cat} and implies abundances of BHs of around
75-100\% (assuming $\dimax = 45^{\circ}$ or $10-100\%$ (for $\Delta i_{\rm max} \le 20^{\circ}$), see Figure \ref{fig:corner_adt_bh}).
As the ULXs in the \cite{Earnshaw_ULX_cat} catalogue have only been observed 2-3
times, there is an observational bias towards alive systems and we underestimate the
true number of transients.  The inferred percentage of  transient to observed systems
is therefore only a lower limit and, as more transients are located, the upper limit
on $\%_{\mathrm{BH}}$ we would infer from Figure \ref{fig:corner_adt_bh} will steadily
push to smaller values.

We also note that the luminosities obtained via sampling the observed population of
\cite{Earnshaw_ULX_cat} are subject to interstellar absorption, whilst the
luminosities obtained from our simulations do not account for this effect.
As such our results are most valid for observations made out of the Galactic
plane and of other galaxies viewed at low inclinations.

Having established in section \ref{sec:lc_dep_on_params} that the relative abundance
of \textit{alive}, \textit{hidden} and \textit{transient} sources may serve to
provide diagnostic information on the quantities describing the underlying population,
we now explore the broad implications for constraining the underlying
nature of the observed ULX population using \erosita.

Following the method described in section \ref{sec:erass_sampling_routine},
we subject the transient light curves to regular sampling, matching the
cadence of \erass, and investigate whether any of our measured quantities,
such as the relative number of transient to
observed sources,
are affected by our input parameters, e.g. the black hole percentage, maximum precessional angle or period
prescription). Figure \ref{fig:erass_evolution_002} shows
three directly observable quantities and their evolution over the course of \erass:
the cumulative number of alive and transient ULXs, and the proportion of transient
to total observed ULXs, for five different black hole ratios (with 1$\sigma$
bounds on the quantities via 10,000 MC iterations). The model parameters used to make Figure \ref{fig:erass_evolution_002} are $Z=0.02$, $d=0.2$, $\dimax = 45^\circ$ (left-hand column)
\& $20^\circ$ (right-hand column), and here we use the Lense-Thirring precession period 
(eqn \ref{eq:Period_wind}).

\subsubsection{Observational predictions for \erass} \label{sec:obs_pred_erass}

\begin{figure}
\centering
\includegraphics[trim={0 0.2cm 0 0}, clip]{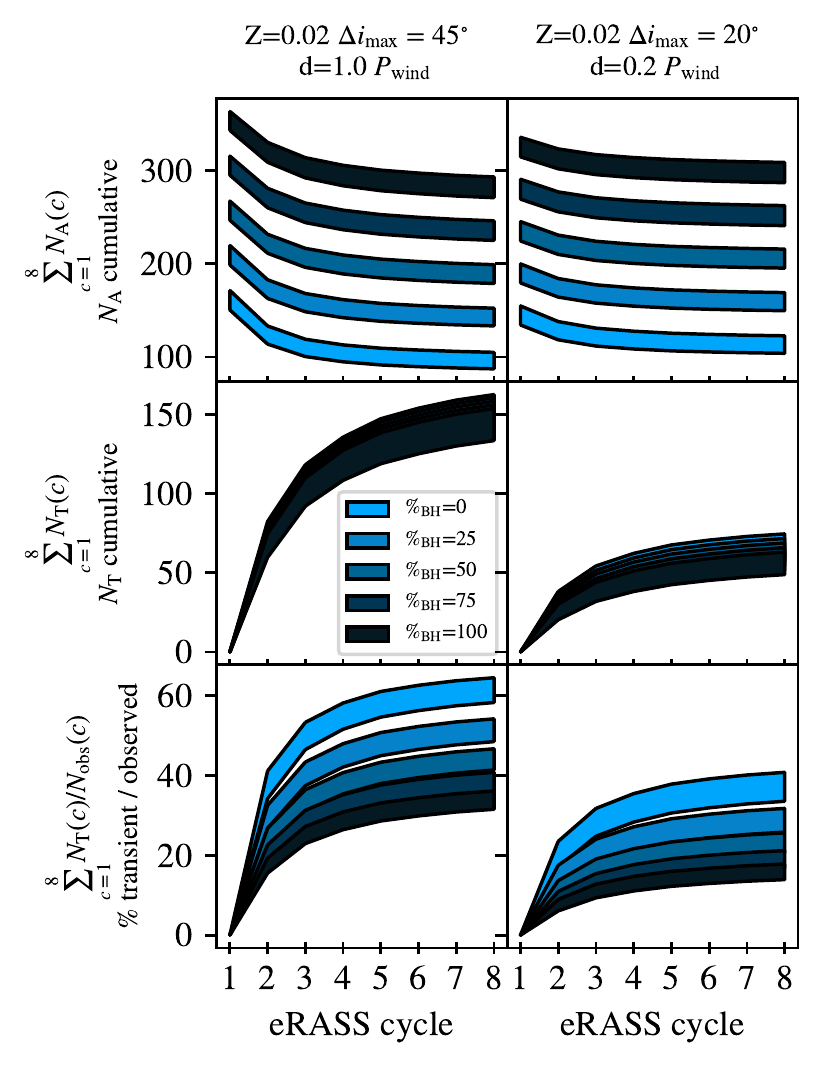}
\caption{
The evolution over \erass \ cycles of three quantities that may be 
directly observed: the number of alive (top row), transient (middle) and
proportion of transient to observed sources (bottom).
Each column shows a different set of model parameters,
while the different colours correspond to the underlying black hole ratio, with darker colours
corresponding to a higher abundance of BHs. The width of the lines indicate the 1$\sigma$ error
regions of the quantities being explored.
}
\label{fig:erass_evolution_002}
\end{figure}

From the first row of Figure \ref{fig:erass_evolution_002}, we observe the
absolute number of \textit{alive} sources detected by \erass \
appears to be sensitive to the underlying black hole ratio, with the number
being positively correlated with the proportion of BHs in the underlying populations.
We also observe that the number of transient sources (middle row) detected by \erass \
is not sensitive to the black hole ratio, i.e. for a given set of parameters ($Z$, $\dimax$, $d$ \& $P$),
the inferred 1$\sigma$ regions overlap. Instead we find that the number of transient sources {\it is}
sensitive to the maximum precessional angle,
with $\dimax = 20^{\circ}$ providing around half the number of transients when compared
to $\dimax = 45^{\circ}$. This may mean that if we have a well-informed prior on the
maximum precessional angle, we may, by considering the relative percentage of
transient to observed sources obtain some indication of the black hole percentage in the 
underlying population.

\begin{figure}
\centering
\includegraphics[trim={0 0.2cm 0 0}, clip]{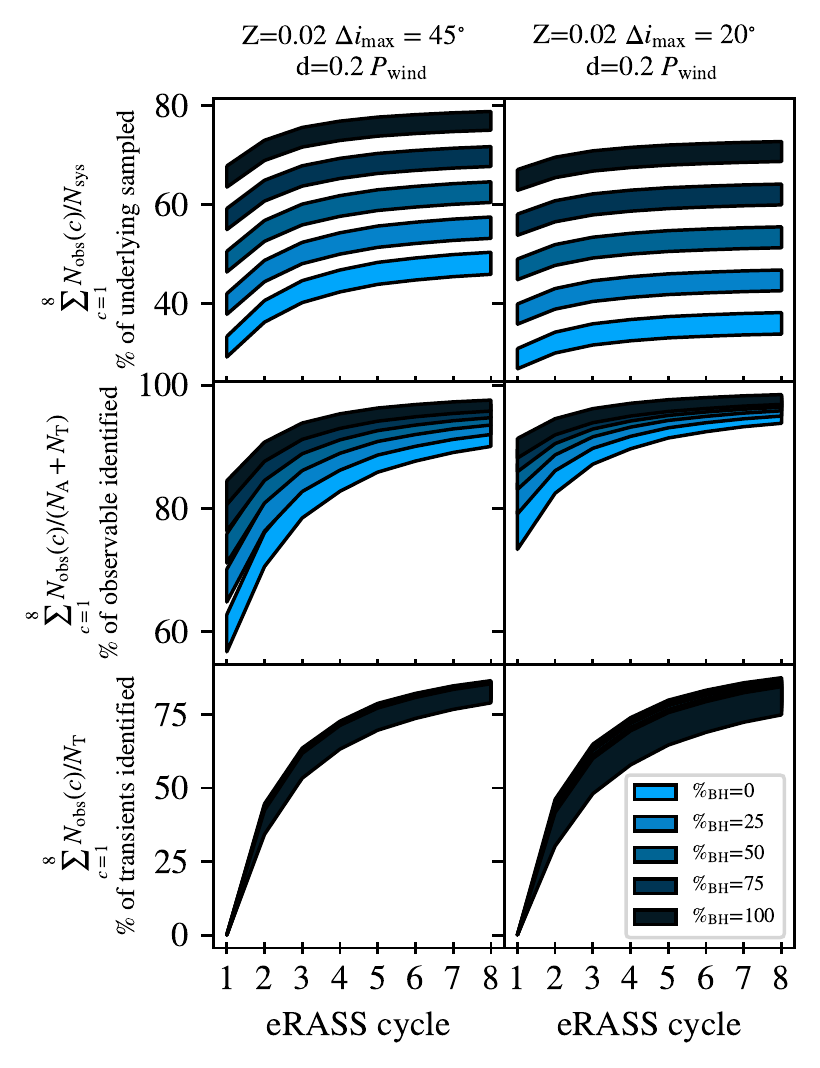}
\caption{
The evolution over \erass \ cycles of the percentage of the underlying
sources sampled (top), the percentage of the potentially observable sources sampled
(middle) and percentage of transient sources sampled (bottom). The width of the lines
indicate the 1$\sigma$ error regions of the quantities being explored.
}
\label{fig:erass_evolution_non_obs_002}
\end{figure}

In Figure \ref{fig:erass_evolution_non_obs_002} we also show three quantities not directly observable
by \erass \ but useful for gaining insight into the performance of the survey:

\begin{itemize}
    \item $\percobssamp$ provides the percentage of the full ULX population sampled by a given \erass \ cycle
    \item $\percobsdet$ which quantifies the percentage of the {\it potentially observable} population which has been observed
    \item $\perctransdet$ which quantifies the percentage of the transient population only which was sampled
\end{itemize}

We will now briefly summarise the effect of each of our model parameters 
on the observed population as seen by \erass.

\begin{itemize}
    \item  {\bf Effect of metallicity}:
    
        Lower metallicity environments are commonly associated with a higher
        abundance of BH systems, as lower metallicity stars
        experience less mass loss than their higher metallicity counterparts
        and are therefore more likely to end up as BHs \citep{2003_Heger_ApJ...591..288H}.
        However, as we are manually specifying the relative
        abundance of BHs in our simulations, the effect of $Z$
        does not strongly correlate with many of our observable
        quantities.

    \item  {\bf Effect of underlying demographic}:
    
        Figure \ref{fig:erass_evolution_002} shows the effect of changing the percentage
        of black holes within the sample, for a given set of model parameters over \erass \ cycles.
        We observe that there is a sizeable increase in the number of alive
        systems for higher abundances of black holes. There is also an essentially constant
        number of transient sources across all black hole ratios. The combination of these last two
        effects means that the percentage of transient to observed sources also shows a
        dependence on the black hole ratio.
        For the set of simulation parameters shown in Figure \ref{fig:erass_evolution_002}, and for a maximum precessional angle of $45^{\circ}$,
        it can be seen that, for a population composed entirely
        of black holes, around $\sim30\%$
        of observed sources may be identified as transient by cycle eight, whilst up to $\sim 60\%$ would be observed as transient for a population composed entirely of neutron stars. For a maximum precessional
        angle of $20^{\circ}$, these values are instead around $\sim20\%$ and $\sim40\%$ respectively.
        
        From Figure \ref{fig:erass_evolution_non_obs_002} we also note that the underlying
        (both full and potentially observable) ULX populations are better sampled for higher black hole percentages in the underlying population.

    \item  {\bf Effect of maximum precessional angle}:
    
        While the Galactic ULX SS433 is well known to have a precessional
        half-angle of $\sim 20^{\circ}$ \citep{1979_Fabian_MNRAS.187P..13F},
        the light curve of NGC 5907 X-1 was described using \ulxlc\ with a
        precessional half-angle of only $\Delta i = 7.30^{+0.13}_{-0.15}$
        \citep{Dauser_2017_ULXLC}. With only two observational constraints (the one for NGC 5907 X-1 naturally being model-dependent),
        the precessional angle, $\Delta i$, remains one of the least constrained free parameters
        in our analysis. As such, we have throughout this work assumed a flat prior, however,
        the physics of the underlying precession mechanism (e.g. in the case of Lense-Thirring precession, the misalignment angle) could plausibly result in
        precessional angles which tend towards the smaller range of values.
        
        
        From Figure \ref{fig:erass_evolution_002}, a maximum precessional angle of
        $\dimax = 45^{\circ}$ roughly halves the absolute number of transients detected in each
        \erass \ cycle while also halving the percentage of transients to total observed systems when
        compared to a maximum precessional angle of $\dimax=20^{\circ}$.
        As seen in Figure \ref{fig:erass_evolution_non_obs_002}, a larger precessional angle results in a higher number of the potentially observable 
        (alive or dead) sources being identified but interestingly results in a lower proportion of
        the entire underlying population being sampled.

    \item  {\bf Effect of the precession prescription}:
        
        Remarkably we find that both the empirical relation of \cite{Townsend_2020} (eq \ref{eq:P_sup})
        and the prediction from the Lense-Thirring model (eq \ref{eq:Period_wind}) produce
        similar results (see Figure \ref{fig:erass_corner_period}).
        This is intriguing as it implies that, regardless of the mechanism, if the disc and wind are
        precessing then we can infer the properties of the underlying sample. Of course,
        should the mechanism be substantially different (e.g. precession of the curtain \citealt{Mushtukov_2017_opt_thick_env_NS}), then this assertion may be invalid.
        
    \item  {\bf Effect of the LMXB duty cycle}:
    
        We find that a lower duty cycle for the LMXB ULX population serves to reduce the
        absolute number of transients detected in each cycle. However, when we consider the
        relative proportion of transients to the total number of observed sources, we find the
        impact of changing the duty cycle to be negligible. 
       
\end{itemize}

\section{Discussion}

The relative proportion of black holes to neutron stars within the observed ULX
population still remains an important unanswered question; of the current
sample of roughly 500 ULXs, around ten are confirmed to have NS accretors, and there
are strong indications that certain objects may harbour black holes (e.g.
\citealt{Cseh_2014_HoII_jets}), but for the vast majority of the population,
the nature of the accretor remains unknown. \cite{Wiktorowicz_2019_obs_vs_tot}
approached this issue by analysing how anisotropic emission of radiation
(geometrical beaming) affects the observed sample of ULXs, finding that, in
regions of constant star formation, the expected number of NS ULXs is higher
than the total number of BH ULXs, however due to the effect of beaming, they
concluded that the total \textit{observed} population was potentially
comparable (cf. \citealt{Middleton_2017_demographics_from_beaming}).  Our work
has built on this by exploring the additional effect of precession of the wind
cone. 

Our simulations have allowed us to construct synthetic XLFs (Figure
\ref{fig:XLF_BH_NS}) and explore the changes resulting from varying the
underlying population demographic. Fitting to only the high luminosity end
(L $\ge 1\times10^{39} \ \ergss$) appears to indicate a range of slopes which are
somewhat steeper than observation (Table \ref{tab:xlf_fit} and Figure \ref{fig:xlf_comparison}) at least where the percentage of black holes in the underlying population are non-zero. It may be that the proportion of black holes is indeed low (as one would expect many more neutron stars than black holes in the intrinsic, underlying population: \citealt{Wiktorowicz_2019_obs_vs_tot}), however there are also several effects which may contribute to differences between simulation and observation. It is important to note that the XLFs we have created from simulation
represent a time-averaged and idealised view of a large population of ULXs, whilst XLFs constructed from single (or from a small number of) observations
instead suffer from a bias towards detecting bright, persistent ULXs rather than transient ULXs (and will also depend on the star formation history of the
target galaxy which we have not accounted for \citealt{2013_Fragos_ApJ...764...41F, Fragos_2013_reionization}). We also note that -- unlike the observational XLFs -- our simulated luminosities
do not assume any absorption; whilst accounting for this effect is complicated (it for instance depends on the unknown spectral shape and local column of a given ULX at a given point in its precessional cycle, e.g. \citealt{2015_Middleton_MNRAS.454.3134M}), this is unlikely to have a major effect
as long as the line-of-sight column is low. Finally, we have assumed a form for the beaming which does not take into account the full complexity of the system, e.g. radial collimation profile, re-processing and outwards advection, all as functions of accretion rate. These complicating effects could potentially bring the highest sources down to lower luminosities, making the XLF steeper.



One of the key results to emerge from our analysis is the indication that a
measure of the relative number of {\it transient} to {\it observed}
ULXs can constrain the nature of the intrinsic population.
Such a result is naturally important as it would allow for a more
concrete understanding of the accreting binary population and related fields
(i.e. studies relying on binary population synthesis, e.g.
\citealt{Fragos_2013_reionization}). However, it is important to consider the
limitations of our approach. We have made the explicit assumption that either
Lense-Thirring torques ($P_{\mathrm{wind}}$) or a different unspecified process
$(P_{\mathrm{sup}}$: \citealt{Townsend_2020}) are the dominant form of variability
on the timescales we are investigating. Whilst Lense-Thirring torques are
certainly unavoidable where the compact object is misaligned (expected in light
of the time required to align the binary -- see \citealt{King_Nixon_2018}),
there are other torques which can dilute or dominate over this effect. These
are discussed at length in \cite{Middleton_2018_Lense_Thirring} but perhaps
most notably we might expect radiation pressure driven warps and precession
\citep{Pringle_1996_warping_disc}, or neutron star dipole precession (see
\citealt{Mushtukov_2017_opt_thick_env_NS}) to occur where the field is very
strong (in the case of the former, the outer disc can be essentially unshielded
for high dipole field NSs, unless the accretion rate is extreme).
We also note that free-body precession may occur as a result of neutron star oblateness
and misalignment of the rotation axis with the axis of symmetry of the star \citep{2001_Jones_MNRAS.324..811J}. This latter effect
has been explored as an alternative origin for the month timescale modulation
seen in ULXs \citep{2020_Vasilopoulos_MNRAS.491.4949V}.

In addition -- and unlike our consideration of the impact of a LMXB duty cycle
-- we have not included the effect of propeller states which occur when neutron
stars are close to spin equilibrium. In such cases, increasing the neutron star
spin by a small amount leads to a period of relative quiescence where emission
from the accretion column and accretion curtain is switched off due to the
centrifugal barrier. If the accretion rate is high or dipole field strength low enough,
then we still expect radiation to emerge from the disc between $r_{\rm sph}$
and the magnetospheric radius, $r_{\rm M}$, which could be substantial (the luminosity then going as $\ln(r_{\rm  sph}/r_{\rm M}$). However, where
the dipole field strength is high or accretion rate low, entering a propellor state could effectively switch
off most of the emission, potentially dropping the source below the empirical ULX threshold.

Throughout this work we have made the assumption that NSs have a low spin
of $a_{*} = 0.01$ while black holes have a maximal spin of $a_{*} = 0.998$. The former is based on the observation of $\sim$1 s periods in ULX pulsars to-date (see \citealt{King_Lasota_2020_PULX_iceberg_emerges} and references in introduction). Naturally, we cannot rule out higher spins for NS systems (as an example, the fastest known spin frequency of a NS at 716Hz
\citealt{2006_Hessels_Sci...311.1901H} would correspond to a maximal spin of $a_{*} = 0.2 - 0.3$ \citealt{2015_Miller_PhR...548....1M}, which would reduce the Lense-Thirring precession timescale accordingly, but the lack of evidence for such spins in ULXs presently limits our ability to explore this. Black hole ULXs may also not be maximally spinning (implying a slower precession period if Lense-Thirring), however, once again we have limited information at this time.

It is interesting to note that around half of the known PULXs appear to be
transient ULXs, with luminosities spanning over a factor of 100
\citep{2019_Song_hunt_for_pulx}. A propeller state has already been reported
in one NS ULX to-date (\citealt{NSULX_Furst_2016}, although the spin evolution
implies the drop in flux is instead driven by obscuration/precession:
\citealt{2021_Furst_A&A...651A..75F}).  \cite{2018_Earnshaw_propeller_MNRAS.476.4272E} have
searched for propeller state ULXs within the entire
\textit{XMM-Newton} 3XMM-DR4 serendipitous source catalogue, identifying five ULXs that
demonstrated long term variability over an order of magnitude in brightness,
while one source (M51 ULX-4) demonstrates an apparent bi-modal flux distribution
that may be consistent with a source undergoing propeller (although this may also
be due to sampling a precessional light curve (e.g. \citealt{Dauser_2017_ULXLC}).
They also note that there are potentially
up to $\sim 200$ sources in the \textit{XMM-Newton} catalogue which may simply
lack a sufficient number of observations using \textit{XMM-Newton} to
reveal their transient nature. Subsequent simulations by the same authors
suggest that \erosita \ may be able to identify 96\% of sources that are
undergoing the propeller effect by cycle 8 of \erass \ (for a duty cycle of
0.5).  This means that if NS ULXs undergoing the propeller effect are present
in a large number within the population, the true number of transient sources in
this paper could be largely underestimated. 

In {\it practice} this means that, without an indication of whether a given source's variability is driven by precession or propeller, the regular observations
taken within \erass \ may lead us to somewhat overestimate the underlying number of transients driven by precession (although this relies on the sample being large and not many sources precessing on very long timescales). As a result, we would tend to over-estimate the abundance of neutron stars in the underlying population. However, if we are able to isolate sources that display precession
(e.g. via fitting of long term light-curves or ruling out the propeller effect), then, given a large enough sample, we would then obtain a {\it lower}
limit on the number of transient (via  precession) to observed sources and a lower limit on the the abundance of neutron stars in the underlying population.

Finally, we have assumed that NSs in our simulations may only reach ULX
luminosities via geometrical beaming, while it is possible
that a drop in the electron scattering cross section due to a high strength
magnetic field, as well as the structure of the accretion column itself could also boost the luminosity (e.g.
\citealt{Basko_Sunyaev_1976, Mushtukov_2017_opt_thick_env_NS}).

\section{Conclusions}

Starting from a synthetic population of binary systems, and using a simple
geometrical model for a precessing cone of emission, we have investigated the
effect precession and beaming might together play on the observed population of
ULXs. We have investigated the effect precession has on the XLF and the
relative numbers of \textit{alive} (persistently $\ge$ 1$\times$10$^{39} \
\ergss$), \textit{transient} (varying across 1$\times$10$^{39} \ \ergss$) and
\textit{hidden} (persistently $<$ 1$\times$10$^{39} \ \ergss$) sources, and, by
factoring in the observational cadence of \erass, we have made
predictions for how well the underlying population may be constrained over the
course of four years of monitoring.

In this paper we propose a novel method for constraining the underlying
demographic within the population of ULXs, as the percentage of ULXs observed
to be transient or observed is sensitive to parameters such as maximum precessional
angle, and crucially to the relative fraction of BHs and NSs in the underlying
population (whilst not sensitive to the duty cycle of LMXB ULXs). This follows from
the fact that -- under the assumptions of geometrical beaming -- populations
containing a higher percentage of BHs are observationally associated with higher
percentages of systems persistently above $10^{39} \ \ergss$ 
and with lower percentages of transient systems, when compared to populations
dominated by NSs.

Determining the underlying ULX demographic presently relies on detecting unambiguous indicators for
the presence of a neutron star such as pulsations or a CRSF. However, it has
been proposed that many NS ULXs with high accretion rates may not exhibit
pulsations \cite{King_2017_Pulsating_ULXs}, that large pulse fractions may be
absent in the presence of strong beaming
\citep{2021_Mushtukov_MNRAS.501.2424M}, and CRSFs may not fall within the
accessible X-ray energy range or may be diluted (see
\citealt{Mushtukov_2017_PULXs_as_magnetars}).  An independent and simple method
to constrain the nature of the underlying population in ULXs such as the one we
have explored here is therefore of value (and joins others such as observing the evolution of
quasi-periodic oscillations, see \citealt{Middleton_2019_Accretion_plane}).

In an initial application of our approach, we have used the
\cite{Earnshaw_ULX_cat} catalogue of ULX and ULX candidates (accepting that this catalogue is incomplete relative to a true flux-limited survey). Finding that $\sim
80\%$ of the catalogue ULXs are always visible, while $\sim 20 \%$ are
transient; this implies a black hole percentage in the underlying population in the
region of $10-100\%$ (for $\Delta i_{\rm max} \le 20^{\circ}$) or $75-100\%$ (for $\Delta i_{\rm max} \le 45^{\circ}$). However, the
number of transients (which we expect to be mostly neutron star ULXs) is likely
to be highly underestimated in such low cadence, pointed observing.
The introduction of \erosita \ and its all sky survey, \erass,
will revolutionise our view of the transient X-ray sky and is optimally placed
to better constrain the underlying demographic of ULXs via this approach. Simulating using two different prescriptions for the precession period:
Lense-Thirring (\citealt{Middleton_2019_Accretion_plane}) and empirical (\citealt{Townsend_2020}), we predict a variety of observational
possibilities for the evolution of the relative numbers of transient to
persistent ULXs over the course of \erass, for a variety of population
characteristics. We conclude that neither prescription for precession
significantly alters our observed view of the ULX population.

We have invoked several simplifications in this work which we will improve upon
in future.  Our model for precession ({\texttt{ULXLC}}) currently does not
account for the energy dependence of the emission; we are developing models
which account for the radial dependence of beaming and which will improve on
the accuracy of our simulations.  We also presently have limited constraints on
the precession angle of the wind cone in ULXs which can have a significant
impact on our predictions; this can be estimated through direct modelling
\citep{Dauser_2017_ULXLC} and, in future, will be developed and applied more
widely to improve our constraints.  Finally, we have not included the effects of
magnetic fields in the neutron star systems in our population; this can have
the effect of changing the X-ray spectrum and beaming but, perhaps more
importantly, can lead to periods of relative quiescence via the propeller
effect \citep{NSULX_Furst_2016, 2018_Earnshaw_propeller_MNRAS.476.4272E} as
well as dipole precession on $\sim$month timescales when the field is very
strong \citep{1980_Lipunov_SvAL....6...14L}.

\section*{Acknowledgements}

NK acknowledges support via STFC studentship
project reference: 2115300.

\section*{Data Availability}
The data obtained from \startrack \ underlying this article are freely accessible at the following urls:\\
\url{https://universeathome.pl/universe/pub/z02_data1.dat}\\
\url{https://universeathome.pl/universe/pub/z002_data1.dat}\\
\url{https://universeathome.pl/universe/pub/z0002_data1.dat}\\


\bibliographystyle{mnras}
\bibliography{references} 




\appendix
\section{Comparison of Period Prescriptions}
\begin{figure*}
    \includegraphics[scale=1.0]{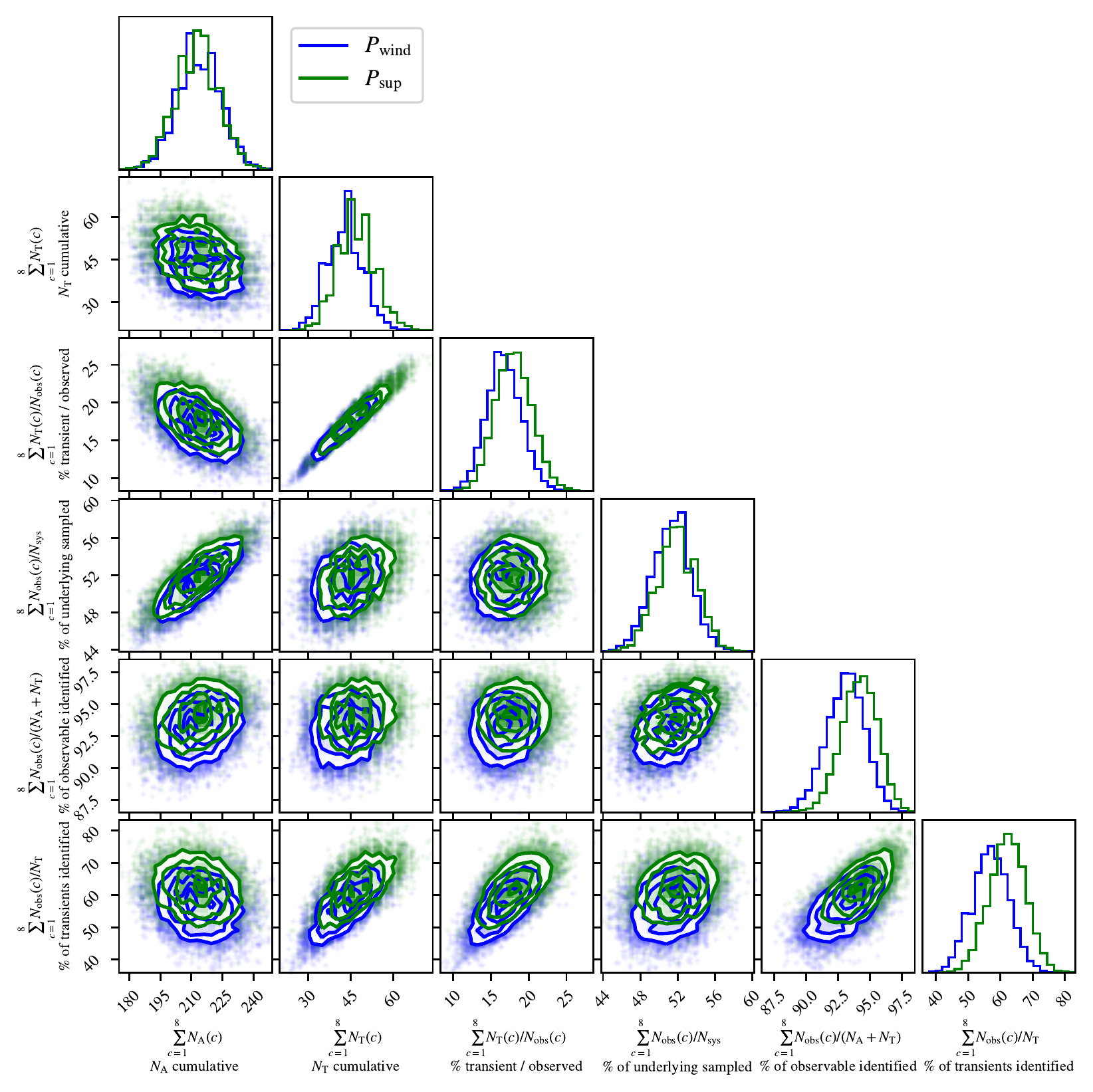}
    \caption{
    Corner plot comparing two precession mechanisms for cycle 3 of \erass \
    showing the minimal impact of using $P_{\mathrm{wind}}$ over $P_{\mathrm{sup}}$.
    Model parameters $Z=0.02$, $\%_{\mathrm{BH}} = 50$, $\dimax$ $d=0.2$.
    } 
    
\label{fig:erass_corner_period}
\end{figure*}


\bsp	
\label{lastpage}
\end{document}